%% file: ICRCFinal.tex
\documentclass[a4paper,11pt]{article}

\usepackage{pos}
\usepackage{subcaption}
\usepackage{lipsum} %package to generate placeholder text in the following
\usepackage{multicol}
\usepackage[left]{lineno}
\usepackage{wrapfig}

\title{Improving Gamma-ray Source Searches with Image Processing}

\ShortTitle{Improving Gamma-ray Source Searches with Image Processing}

\author{HAWC Collaboration \\{\normalsize \normalfont(a complete list of authors can be found at the end of the proceedings)}\\}

\emailAdd{rbabu@icecube.wisc.edu}
\emailAdd{itomura@msu.edu}
\emailAdd{herzogia@msu.edu}
\emailAdd{salaza82@msu.edu}
\emailAdd{nisamehr@msu.edu}
\abstract{
The High Altitude Water Cherenkov (HAWC) observatory surveys the very-high-energy sky in the energy range from 300 GeV to greater than 100 TeV. Its wide field of view makes it particularly suitable for studying large, extended regions of emission that often contain multiple point and diffuse sources of gamma rays. Existing blind search methods to detect sources in HAWC data use a computationally expensive, iterative source fitting algorithm that can take several days to scan a few degree region in the sky. In this work, we adopt a new approach to speed up the identification of sources using an image processing pipeline. Using image processing filters, and a blob finder algorithm, based on the Determinant of a Gaussian, this pipeline seeds sources accurately up to 300 times faster than the current HAWC source search pipeline. The pipeline's output are then passed onto a global multi-threaded likelihood fitter for accurate source localization. We present the performance of the improved pipeline and discuss prospects for future applications on other astrophysical datasets. 

\vspace{4mm}

{\bfseries Corresponding authors:}
Rishi Babu$^{1}$, 
Palmer Wentworth$^{1}$, 
Ian Herzog $^{1*}$, 
Dan Salazar$^{1}$,
Mehr Un Nisa $^{1}$\\
{$^{1}$ \itshape Department of Physics \& Astronomy, Michigan State University, Lansing, MI, USA}\\
$^*$ Presenter
}

\input{ICRCdetails.tex}

\begin{document}

\maketitle

\section{Introduction}\label{sec1}

The analysis of HAWC data is done using a likeihood-based systematic multi-source search procedure utilizing the threeML software package and the HAL Plugin \citep{3ml}\citep{hal}. This approach was inspired by the extended source search described in the Fermi-LAT Extended Source Search Catalog \citep{fges}. The pipeline to search for point sources and extended sources within the region of interest (ROI) is described in \citep{1809}. This procedure is computationally intensive and often requires multiple days to produce a final model for a region. Often this approach misses fainter sources, which need to be added manually later. This blind search could be complemented by providing a method to seed source locations, which can be assessed and evaluated by likelihood analysis, thereby significantly reducing the computational time and resources spent on finding a model. 

In this work, we present the results of an image processing algorithm, using the \texttt{scikit-image} \cite{skimage} python-package, to seed sources from a gamma-ray significance image. Additionally, we show the performance metrics of the image processing pipeline with 60 simulations and compare the speed of the pipeline with conventional likelihood source search method. Finally, we discuss the applications of the image processing algorithm in the context of source seeding in astronomical images, also by extending the application to Fermi-LAT, and potentially with future public LHAASO and SWGO data.

\begin{figure} [ht!]
    \centering
    \includegraphics[width=0.8\textwidth]{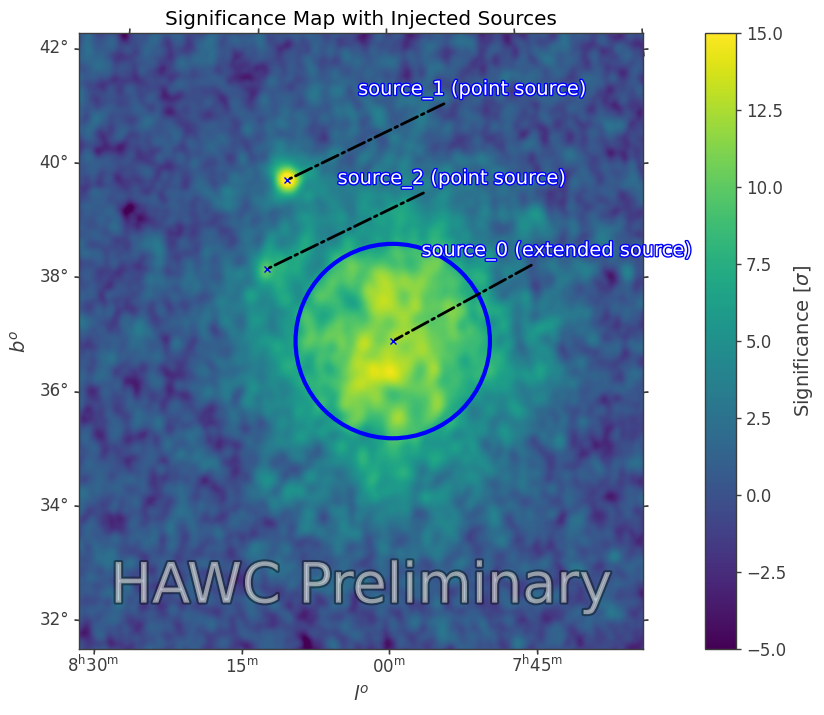}
    \caption{HAWC Significance Sky Map showing the location and morphology of three different simulated sources, marked with their labels and extensions (in galactic coordinates).}
    \label{fig:skymap}
\end{figure}

%%%%%%%%%%
%\begin{centering}{r}{\textwidth}
%    \centering
%    \includegraphics[width=0.38\textwidth]{ICRC2025templates/plots/Map.png}
%    \caption{Normalized azimuth distributions for neutrino-induced cascades from MC simulations only. Solid lines compare the EventGenerator predictions using the FTP-v3 ice model (red) and the BFR ice model (green), while the filled histogram (light blue) shows the same FTP-v3 prediction processed through the PreferredFit reconstruction.}
%    \label{fig:ice_modeling}
%\end{centering}%

\section{Source Detection Pipeline}
To analyze and seed sources in a region, the first step is to determine the region of interest (ROI) of the source in question. The ROI of the image is chosen to encompass a large region ($\approx 5^0 \times 5^0 $) around the potential sources. Once the ROI is found and the image is selected, it is then passed on to the image processing pipeline. The image processing pipeline is described below: 
\subsection{Image Preprocessing}

The significance image is pre-processed to remove significant under-fluctuations. Significance maps show severe under-fluctuations, in regions where there is not enough background statistics and near the edges of the field of view (FOV) of the detector, as negative significance. When the significance is $ < \sim -5 \sigma$, during the image contrast normalization and blob detection steps (described in the following sections), background fluctuations starts to show up as fake source seeds (noise seeds). To eliminate these noise seeds, the original significance map is clipped to a minimum value of -5 $\sigma$, the resulting image is shown in Fig.(\ref{fig:skymap}). This is done to reduce the noise in the images, while still keeping a large dynamic range for the pixel intensities for further processing. The clipped image is then scanned for peaks with values $\geq 5 \sigma$ intensity, which is our detection threshold. If there are peaks found in the map, then the pipeline proceeds to the next step. 

\subsection{Image Normalization}
The clipped image shows a large dynamic range which is based on the ROI of the region in analysis, as regions with fainter sources look different from those with very bright sources. To produce consistency in the dynamic range between different regions of the data and to reduce the bias of finding brighter sources, the clipped image is first normalized to values between [0, 1]. This is an essential technique for spatial-based image enhancement, as it adjusts the image contrast to bring out the fainter sources in the images. 

\subsection{Feature Enhancement using Difference of Gaussian method}
The normalized image, does not show the presence of any underlying features buried in it. To look at the underlying features in the normalized image, we use a feature enhancement algorithm called the Difference of Gaussians (DoG)\cite{dogcite}\cite{dogpaper}. 

\begin{figure}[ht!]
    \centering
    \begin{subfigure}[t]{0.5\textwidth}
        \centering
        \includegraphics[height=1.8in]{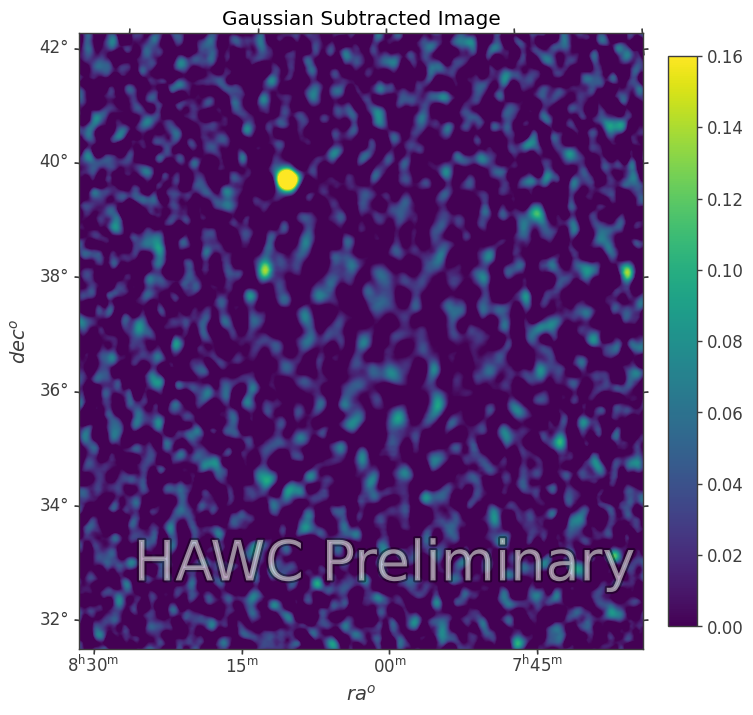}
        \caption{Residual map after subtracting the $0.2^0$ Gaussian convolved kernel image from the normalized image}\label{gresmap}
    \end{subfigure}%
    ~ 
    \begin{subfigure}[t]{0.5\textwidth}
        \centering
        \includegraphics[height=1.8in]{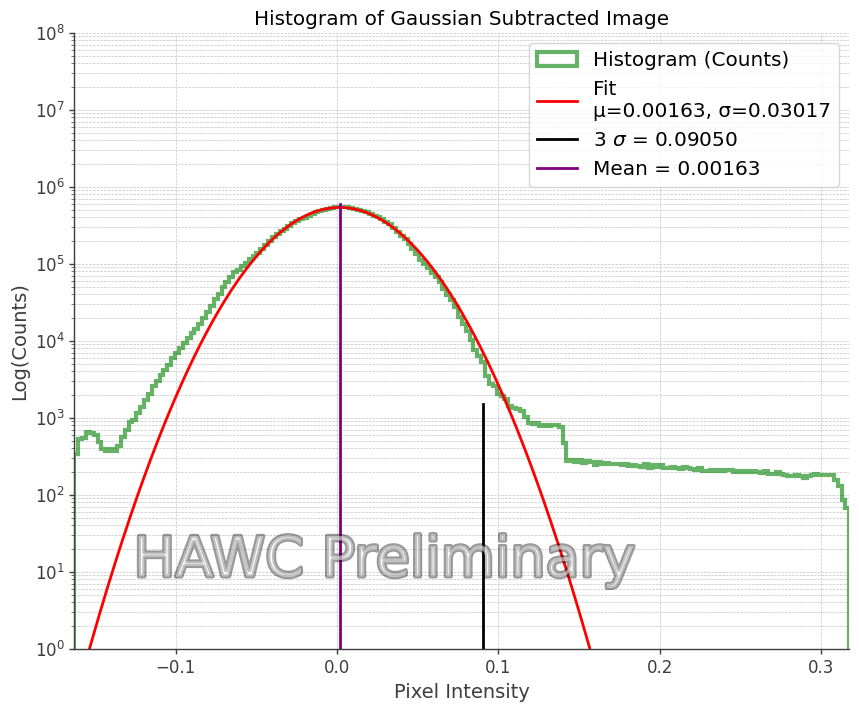}
        \caption{Intensity Histogram of the DoG residual map. The pixel intensitiy curve are show in green, the gaussian fit to the curve is shown in red. The black line shows the $3\sigma$ deviation from the gaussian fit.}\label{rgauss}
    \end{subfigure}
    \caption{Difference of Gaussian (DoG) Algorithm}\label{gbdt_performance_metrics}
\end{figure}

The normalized image is convolved with a Gaussian kernel with a smearing radius, equal to the point spread function (PSF) value, averaged over different bins, of the detector corresponding to the declination of the region in analysis. Blurring the normalized image with the Gaussian kernels suppresses high-frequency spatial information such as sharp peaks, sharp edges, fine boundaries.. etc and enhances low-frequency spatial information such as gradual change in intensities. The Gaussian convolved image is then subtracted from the normalized image to produce a "DoG residual image". The resulting image preserves the high frequency components which when examined corresponds to the sources in the images. Point sources and smaller extended sources appear as sharper peaks, whereas large extended sources are removed. From Fig.(\ref{gresmap}), it seen that there are 2 main peaks, which correspond to the point sources in the original map of the simulated sources seen in Fig.(\ref{fig:skymap}), and the large extended source from the original map is removed. 
Therefore, the DoG algorithm acts as a spatial bandpass filter, allowing high frequency components to pop up. A smearing radius of $0.2^0$ is used in the simulation plots shown in the proceeding. The intensity histogram of the DoG residual image is calculated as shown in Fig.(\ref{rgauss}).The intensity histogram (green) resembles a zero-centered Gaussian distribution with a skewed tail toward positive values. Therefore, the intensity data is fit with a gaussian function (red), and we calculate the $3\sigma$ deviation of the gaussian intensity, which is used for intensity thresholding in the next step.

\subsection{Blob Detection and Intensity Thresholding}
We employ a blob detection algorithm, which is based on the DoG approach to look for blobs in the images \cite{dogpaper}. The blob detection algorithm works by building an array of images, in which the  DoG residual image is blurred with increasing smearing radii and the difference between two consecutive images are stored in a 3D array. The algorithm scans each of these images in the 3D array, where it looks for high intensity at the location of each pixel. For each pixel scanned in a single image, it performs a secondary scan of the 8 neighboring pixels. The the pixel is evaluated by comparing its intensity to the intensity of the surrounding pixels and determines if its a local maxima/minima, called as the local extrema. If the pixel is local extrema, then the pixel is compared to its corresponding pixel in  all the images from the 3D array, and if the algorithm shows a local extrema in multiple blurred scale images, then it tags the pixel as a blob and the (x,y) pixel position and scale size is returned. 

The DoG blob finder returns multiple detected blobs, both real and fake, depending on the local extrema and fluctuations near the boundary of the images. To remove the fake blobs, we apply a boundary cut and intensity thresholding steps.

A boundary cut is applied where blobs detected $1^0$ from the edges are removed. Then, an internal intensity thresholding algorithm rejects the detected blobs, whose local extrema are less than the $3\sigma$ deviation from the gaussian fit to the pixel intensity histogram of the DoG residual map. A second thresholding algorithm is applied, where the significance values of the blobs from the input significance map are compared. If the significance value inside the blobs is less than 5 sigma, it is removed from the list of detected blobs.

After filtering the blobs with the intensity thresholds, a source seed list is produced. The seeds in the sources are classified into 2: (a) If the scale of the blob is less than $0.15^0$, then it is classified as a point-like source (b) If the scale of the blob is greather than $0.15^0$, then it is classified as an extended-like source. The seed list is passed then passed onto a astropy model file and the sources are evaluated using threeML+HAL Plugin.

\section{Performance}
We run the pipeline on 60 sets of simulations with a total of 123 sources: (i) 66 point sources (ii) 25:  $< 0.5^0$ extended sources (iii) 33: $> 0.8^0$ extended diffuse sources. We recover 65 point sources, 24 of $< 0.5^0$ extended sources and 1 of $> 0.8^0$ extended diffuse sources. We see that the algorithm has a 10\% false positive rate, most of it coming from detecting very bright $> 0.8^0$ extended sources as point-like. The cpu wall time to run the image processing pipeline, over 60 simulations sets was less than 26 hrs, whereas the conventional blind search using likelihood fitting algorithms took over 2 months. The image processing algorithm achieves a $~ 60\times$ speedup compared to the conventional method, corresponding to an approximate performance gain of 98.33\%.
\begin{figure} [ht!]
    \centering
    \includegraphics[width=0.8\textwidth]{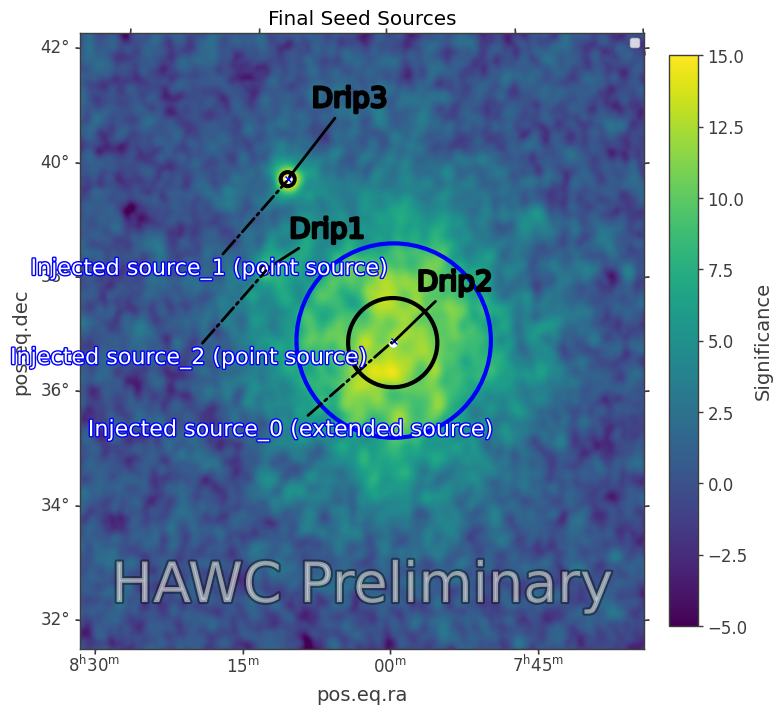}
    \caption{The source seeds detected from the image processing pipeline, along with the simulated source locations.}
    \label{seedper}
\end{figure}

The output seeds, referred to as Drips, generated by the image processing pipeline on the simulation dataset used in this study are shown in Fig.(\ref{seedper}). We see that for the simulation dataset used, the image processing pipeline is fully capable of recovering the simulated sources.

\section{Conclusion}\label{sec x}
We explore the use of modern image processing algorithms for astronomical source detection, demonstrating significantly faster source seeding compared to traditional blind likelihood-based methods. The algorithm presented here achieves >98\% accuracy in identifying point sources and >95\% accuracy for extended sources, while offering a computational speedup of approximately 98\% relative to conventional approaches. The proposed pipeline can be seamlessly extended to source-seeding applications in datasets from instruments like Fermi-LAT, as well as upcoming publicly available data from SWGO and LHAASO, enabling the full utilization of image-based analysis methodologies.

% Bibtex references:
\bibliographystyle{JHEP}
\bibliography{references.bib}

\clearpage

%The following list of authors, affiliations and funding agencies will be updated at the day of submission. The following template is a placeholder generated via https://authorlist.icecube.wisc.edu/icecube on May 17, 2025 and will be updated.
\input{authorlist_hawc.tex}

\end{document}

%% file: ICRCdetails.tex
\ConferenceLogo{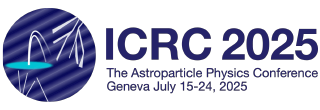}

\FullConference{39th International Cosmic Ray Conference (ICRC2025)\\
 15–24 July 2025\\
Geneva, Switzerland\\}

%% file: authorlist_hawc.tex
\clearpage
\section*{Full Author List: \ HAWC Collaboration}
\scriptsize
\noindent
%first.author$^1$, 
%second.author$^2$, 
%third.author$^3$ % .... more names
%and 
%last.author$^{n}$ \\
%
%\noindent
%$^1$first.affiliation.
%$^2$second.affiliation. % .... more affiliation
%$^{m}$last.affiliation.
\vskip2cm
\noindent

R. Alfaro$^{1}$,
C. Alvarez$^{2}$,
A. Andrés$^{3}$,
E. Anita-Rangel$^{3}$,
M. Araya$^{4}$,
J.C. Arteaga-Velázquez$^{5}$,
D. Avila Rojas$^{3}$,
H.A. Ayala Solares$^{6}$,
R. Babu$^{7}$,
P. Bangale$^{8}$,
E. Belmont-Moreno$^{1}$,
A. Bernal$^{3}$,
K.S. Caballero-Mora$^{2}$,
T. Capistrán$^{9}$,
A. Carramiñana$^{10}$,
F. Carreón$^{3}$,
S. Casanova$^{11}$,
S. Coutiño de León$^{12}$,
E. De la Fuente$^{13}$,
D. Depaoli$^{14}$,
P. Desiati$^{12}$,
N. Di Lalla$^{15}$,
R. Diaz Hernandez$^{10}$,
B.L. Dingus$^{16}$,
M.A. DuVernois$^{12}$,
J.C. Díaz-Vélez$^{12}$,
K. Engel$^{17}$,
T. Ergin$^{7}$,
C. Espinoza$^{1}$,
K. Fang$^{12}$,
N. Fraija$^{3}$,
S. Fraija$^{3}$,
J.A. García-González$^{18}$,
F. Garfias$^{3}$,
N. Ghosh$^{19}$,
A. Gonzalez Muñoz$^{1}$,
M.M. González$^{3}$,
J.A. Goodman$^{17}$,
S. Groetsch$^{19}$,
J. Gyeong$^{20}$,
J.P. Harding$^{16}$,
S. Hernández-Cadena$^{21}$,
I. Herzog$^{7}$,
D. Huang$^{17}$,
P. Hüntemeyer$^{19}$,
A. Iriarte$^{3}$,
S. Kaufmann$^{22}$,
D. Kieda$^{23}$,
K. Leavitt$^{19}$,
H. León Vargas$^{1}$,
J.T. Linnemann$^{7}$,
A.L. Longinotti$^{3}$,
G. Luis-Raya$^{22}$,
K. Malone$^{16}$,
O. Martinez$^{24}$,
J. Martínez-Castro$^{25}$,
H. Martínez-Huerta$^{30}$,
J.A. Matthews$^{26}$,
P. Miranda-Romagnoli$^{27}$,
P.E. Mirón-Enriquez$^{3}$,
J.A. Montes$^{3}$,
J.A. Morales-Soto$^{5}$,
M. Mostafá$^{8}$,
M. Najafi$^{19}$,
L. Nellen$^{28}$,
M.U. Nisa$^{7}$,
N. Omodei$^{15}$,
E. Ponce$^{24}$,
Y. Pérez Araujo$^{1}$,
E.G. Pérez-Pérez$^{22}$,
Q. Remy$^{14}$,
C.D. Rho$^{20}$,
D. Rosa-González$^{10}$,
M. Roth$^{16}$,
H. Salazar$^{24}$,
D. Salazar-Gallegos$^{7}$,
A. Sandoval$^{1}$,
M. Schneider$^{1}$,
G. Schwefer$^{14}$,
J. Serna-Franco$^{1}$,
A.J. Smith$^{17}$
Y. Son$^{29}$,
R.W. Springer$^{23}$,
O. Tibolla$^{22}$,
K. Tollefson$^{7}$,
I. Torres$^{10}$,
R. Torres-Escobedo$^{21}$,
R. Turner$^{19}$,
E. Varela$^{24}$,
L. Villaseñor$^{24}$,
X. Wang$^{19}$,
Z. Wang$^{17}$,
I.J. Watson$^{29}$,
H. Wu$^{12}$,
S. Yu$^{6}$,
S. Yun-Cárcamo$^{17}$,
H. Zhou$^{21}$,

\vskip2cm
\noindent

$^{1}$Instituto de F\'{i}sica, Universidad Nacional Autónoma de México, Ciudad de Mexico, Mexico,
$^{2}$Universidad Autónoma de Chiapas, Tuxtla Gutiérrez, Chiapas, México,
$^{3}$Instituto de Astronom\'{i}a, Universidad Nacional Autónoma de México, Ciudad de Mexico, Mexico,
$^{4}$Universidad de Costa Rica, San José 2060, Costa Rica,
$^{5}$Universidad Michoacana de San Nicolás de Hidalgo, Morelia, Mexico,
$^{6}$Department of Physics, Pennsylvania State University, University Park, PA, USA,
$^{7}$Department of Physics and Astronomy, Michigan State University, East Lansing, MI, USA,
$^{8}$Temple University, Department of Physics, 1925 N. 12th Street, Philadelphia, PA 19122, USA,
$^{9}$Universita degli Studi di Torino, I-10125 Torino, Italy,
$^{10}$Instituto Nacional de Astrof\'{i}sica, Óptica y Electrónica, Puebla, Mexico,
$^{11}$Institute of Nuclear Physics Polish Academy of Sciences, PL-31342 11, Krakow, Poland,
$^{12}$Dept. of Physics and Wisconsin IceCube Particle Astrophysics Center, University of Wisconsin{\textemdash}Madison, Madison, WI, USA,
$^{13}$Departamento de F\'{i}sica, Centro Universitario de Ciencias Exactase Ingenierias, Universidad de Guadalajara, Guadalajara, Mexico, 
$^{14}$Max-Planck Institute for Nuclear Physics, 69117 Heidelberg, Germany,
$^{15}$Department of Physics, Stanford University: Stanford, CA 94305–4060, USA,
$^{16}$Los Alamos National Laboratory, Los Alamos, NM, USA,
$^{17}$Department of Physics, University of Maryland, College Park, MD, USA,
$^{18}$Tecnologico de Monterrey, Escuela de Ingenier\'{i}a y Ciencias, Ave. Eugenio Garza Sada 2501, Monterrey, N.L., Mexico, 64849,
$^{19}$Department of Physics, Michigan Technological University, Houghton, MI, USA,
$^{20}$Department of Physics, Sungkyunkwan University, Suwon 16419, South Korea,
$^{21}$Tsung-Dao Lee Institute \& School of Physics and Astronomy, Shanghai Jiao Tong University, 800 Dongchuan Rd, Shanghai, SH 200240, China,
$^{22}$Universidad Politecnica de Pachuca, Pachuca, Hgo, Mexico,
$^{23}$Department of Physics and Astronomy, University of Utah, Salt Lake City, UT, USA, 
$^{24}$Facultad de Ciencias F\'{i}sico Matemáticas, Benemérita Universidad Autónoma de Puebla, Puebla, Mexico, 
$^{25}$Centro de Investigaci\'on en Computaci\'on, Instituto Polit\'ecnico Nacional, M\'exico City, M\'exico,
$^{26}$Dept of Physics and Astronomy, University of New Mexico, Albuquerque, NM, USA,
$^{27}$Universidad Autónoma del Estado de Hidalgo, Pachuca, Mexico,
$^{28}$Instituto de Ciencias Nucleares, Universidad Nacional Autónoma de Mexico, Ciudad de Mexico, Mexico, 
$^{29}$University of Seoul, Seoul, Rep. of Korea,
$^{30}$Departamento de Física y Matemáticas, Universidad de Monterrey, Av.~Morones Prieto 4500, 66238, San Pedro Garza Garcia NL, Mexico